\documentclass[article,final,moreauthors]{Definitions/mdpi}

\firstpage{1} 
\pubvolume{1}
\issuenum{1}
\articlenumber{0}
\pubyear{2026}
\copyrightyear{2026}

\datereceived{ } 

\dateaccepted{ } 
\datepublished{ } 
\usepackage{comment}

\Title{How long is a photon?}

\newcommand{\m}{p_{(\Delta)}}

\Author{Sky Nelson-Isaacs}
\address{%
Sky Nelson-Isaacs
}

\corres{Correspondence: Sky Nelson-Isaacs, theskyband@gmail.com}

\abstract{A photon has the strange property that it travels along a null interval with zero proper time, hence its emission and absorption points are, in a sense, a singular event. For their part, the emission and absorption processes have a finite duration which determines the spectral content of the photon. The question in the title points to the conceptual difficulty in `stitching' together these three intervals, which all correspond to the same physical entity but have fundamentally incompatible temporal descriptions. A previously developed model for quantum wavefunction propagation solves this apparent dilemma by formulating the life cycle of a photon in terms of recursive Fourier transforms on dual spaces. Because it doesn't treat space as dependent on time, and it is built on the Fourier transform whose phase is an invariant in special relativity as well as the foundation of the uncertainty principle, the model is a promising candidate framework for making quantum mechanics compatible with relativity. The model proposes to serve as an improved Klein-Gordon equation, or relativistically compatible replacement for Schr\"{o}dinger's equation.}

\keyword{klein-Gordon equation; relativistic Schr\"{o}dinger equation; quantum wavefunction propagation; scalar diffraction theory; spectral theory; quantum gravity; holographic spacetime intervals; non-locality; Fourier optics; quantum measurement; single-photon generation} 

\renewcommand{\linenumbers}{}
\begin{document}


\section{Introduction}

The question in the title highlights a gap in our understanding of time. Photons are clearly affected by the duration of their creation and absorption processes, as evidenced by the spectrum they carry, but they are not affected by the time of travel between these two points. In effect, these are a single event in Virtual contact, per Lewis.\cite{LEWIS1926-2} this well known property of light Bridges both special relativity (light propagates along a null interval) and quantum theory (light is generated by atomic electronic transitions), yet is glossed over without much ado. One reason why it may be looked over is that the frame of a photon isn't valid in relativity theory, but clearly causality is communicated across distances, and so the question of how the photon is affected by each interval is a more rigorous approach and highlights the existence of a conundrum: It seems odd to `stitch together' a time-based emission with a time-free propagation and then a time-based absorption in order to construct the full description of a photons duration. 

The solution to this stated problem is hidden in the dual properties of spacetime and its frequency conjugate. By placing these at the foundation of our theory, we can understand the different regions of existence of the photon to be described best in alternate spaces. Then the connection between them is the well-known Fourier transform. We can't actually answer the question ``how long is a photon?'' because it is found that different aspects of the photon's evolution happen in each of the two conjugate domains, which cannot be compared directly. However, we come away with a model that has other benefits, such as relativistic invariance and compatibility with quantum theory.

\subsection{Phase mechanics in digital signals\label{sec:digital_signals}}
To understand the key element that allows for treating all parameters equally can be illustrated with a thought experiment. In digital image processing, visual features can be predictably moved in horizontal or vertical position on a screen by applying the right mask which alters the phase factors of every frequency component appropriately. By applying a single on the original scene, it is discretely altered in a coherent way. This procedure works for the time dimension as well: a piece of music can be converted to the frequency domain and particular sonic features can me translated in time simply through a \textit{static} application of phase.

By treating both space and time as independent, unobservable parameters, this provides a clue that we can modify not only the momentary spatial and temporal locations but also on-going motion of quantum mechanical wavepackets by simply layering on a phase mask. This is an alluring prospect, pulling back the veil on continuous time evolution, suggesting that with the right phase map, any arbitrary spacetime configuration can be manufactured!

\subsection{Background and overview}
The model presented here is can be summarized in the following steps:
\begin{enumerate}
    \item Start with a wave distribution where space and time variables are treated equally, so nothing evolves on the map.
    \item Perform a Fourier transform, bringing the wave distribution into k space.
    \item Evolve every frequency with the appropriate phase to propagate the signal
    \item Perform a Fourier transform again, bringing the wave distribution back into the spacetime domain.
    \item Evolve every spatial frequency in this domain with the appropriate phase to propagate the signal
    \item Repeat these steps
\end{enumerate}

What does this accomplish? Half of the cycle is identical to Fourier optics in which signals are propagated in the frequency domain without a time variable, Shifting the space-time location of the packet. The other half of the cycle is the analogous process in the conjugate domain: instead of shifting the space-time location of the packet, it shifts the momentum space location of the packet. Together, all evolution is accounted for, and by choosing phase functions appropriately, basic kinematics can be replicated via discrete applications of the Fourier transform.

The model has the following features:
\begin{itemize}
    \item Is relativistically invariant
    \item Describes curvature of spacetime
    \item Reproduces dispersion relations associated with mass, as transverse spatial frequency modes of the null interval which require travel at group velocity < c
    \item Accounts for emission and absorption as time-reversal process \cite{Thomas2026}    
\end{itemize}

\subsubsection{Related formalisms}
This approach to Quantum propagation has similarities with a number of well-established formalisms. Principal among these is the lagrangian formulation of Dynamics, in which the entire phase space is considered when determining the allowable trajectories of a system. The result is a invariant path-as-a-whole which defines motion through space with respect to time, but doesn't actually evolve time in order to predict motion. This is done by minimizing the action on a path, and the action minimized is a phase built from the potentials and positions of the parts of the system. In lagrangian Dynamics, the action functional is integrated along a timeline and the coordinates and velocities are still considered to be functions of the independent time coordinate. The theory presented here is the logical extension of the Lagrangian effort to fully liberate physics from asymmetries in space and time using the same tool (phase) while fully abstracting away from objective measurements of time. The key insight not available during Lagrange's time was the development of holography and quantum optics, which provide a clear bridge from Lagrange's work to that presented here.

Some modern variations on quantum theory have close similarities.

In the early days of quantum theory, Wheeler and Feynman developed a `field-free' approach to quantum electrodynamics in which the symmetry between past and future states is preserved by identifying the inner product which produces observables $\psi^* \psi$ to be composed of a retarded wave from the past and an advanced wave from the future.\cite{WheelerFeynman1945} This is motivated by the curious fact that the necessary complex conjugation in the basic formalism (for instance, the Copenhagen interpretation formalized by von Neumann \cite{vonNeuman1932}) can be considered as setting the time evolution variable to negative values, if interpreted literally. In their theory, absorbers in the future must exist in order for emitters in the past to emit their radiation in the first place. This theory nicely illustrates the difference between Newtonian time's oblivious march into the future and the more subtle notion developed for instance in the path integral that the whole path exists before it happens, and the present can't happen without agreement from the future. 

The Wheeler-Feynman approach is a statistical model which does not pair individual past events with individual future events.
The transactional interpretation by John Cramer \cite{cramer1986transactional} extends the Wheeler-Feynman theory to specific ``handshakes'' between emitters and absorbers.

The two-state vector formalism constructs paths-as-a-whole through time by treating the bra and ket in Dirac notation as initial and final states with arbitrary time evolution happening in between. The measurables are shown to be selectable from proper choice of the final state, indicating not only a forward in time causality but also a reverse-time pull from the future, like Plato's final cause. Interesting predictions can be made on the measurable known as the ``weak value'' of a system, but the process' dependence on post-selecting a subset of the ensemble of experimental runs, whether this constitutes a backwards-in-time effect or not is obscured.

The model used to examine the question in the title and perform evolution of systems was developed by the author in prior work, \cite{NELSONISAACS2020} \cite{NELSONISAACS2024} \cite{NelsonIsaacs2025}.

\section{Formalism}
The fundamental structure in this model is a $3+1$-dimensional spacetime distribution, and a conjugate $3+1$-dimensional frequency distribution. Crucially, all eight of these parameters are independent; time is not the central ticking clock for space.\cite{NELSONISAACS2020} These are introduced via two postulates.

Postulate 1. The state of a system is completely characterized by wave distributions $\Psi(x^\mu)$ and $\tilde{\Psi}(k^\mu)$ in complex-valued dual 3+1D spaces parameterized by $x$,$y$,$z$,$t$ and $k_x$,$k_y$,$k_z$,$\omega$ respectively (i.e. replaces the QM wavefunction,  $\Psi(x(t))$), which is explicitly time dependent). These 3+1D parameter distributions evolve via convolution with a phase map in each domain,\cite{NELSONISAACS2020} or using the convolution theorem to rewrite as recursive Fourier transform operations, (written without loss of generality with one spatial and one temporal parameter, 1+1D)
\begin{align}
    \Psi'(x,t) = \mathcal{F}^{-1}\left\{\mathcal{F}\{e^{(iS(x,t))}\}\,\mathcal{F}\{\Psi_1(x,t)\}\right\}
    \label{eqn:RFT_PROPAGATE_X}
\end{align}
\begin{align}
    \tilde{\Psi}'(k,\omega) = \mathcal{F}\left\{\mathcal{F}^{-1}\{e^{-i\tilde{S}(k,\omega)}\}\,\mathcal{F}^{-1}\{\tilde{\Psi}_1(k,\omega)\}\right\}.
    \label{eqn:RFT_PROPAGATE_K}
\end{align}
where $\Psi_1$ is an initial 1+1D signal, ``tilde'' indicates the distribution in $k^\mu$-space, and $\mathcal{F}$ is the Fourier transform (the convolution theorem was used). Note that the propagation involves both forward and inverse Fourier transforms.
    
Postulate 2. Together, the dual spaces comprise a complete description of the dynamical interactions between systems, obtained through (parity inverting) convolution with a phase map in either of the spaces. Again writing with the help of the convolution theorem,
\begin{align}
    \tilde{\Psi}(k,\omega)=\mathcal{F}\{e^{iS(x,t)}\Psi_1(x,t)\},
    \label{eq:RFT_LENS_X}
\end{align}
\begin{align}
    \Psi'(-x,-t)=\mathcal{F}\{e^{-i\tilde{S}(k,\omega)}\tilde{\Psi}(k,\omega)\},
    \label{eq:RFT_LENS_K}
\end{align}
involving repeated \textit{forward} Fourier transforms.
\footnote{It was shown in \cite{NELSONISAACS2020} that these equations can be reduced to the generic Schrondinger equation.}
Compared to the propagation expressions, eqns. $\ref{eqn:RFT_PROPAGATE_X}$-$\ref{eqn:RFT_PROPAGATE_K}$, the interaction expressions invert the signal through the repeated application of forward transforms (no inverse transforms). This results in the reversal of the signal ordering.

\subsection{What is an interaction?}
Crucially, this is a theory of interactions, where field interactions are analogous to optical images passing through a lens onto the image plane. Because evolution happens via a Fourier transform operation, it is a discrete step, compared to the Schr\"{o}dinger equation which can be integrated for a finite length of time. Interactions here are discrete.

Additionally, we will find that the tools used to evolve a system, the phase distributions or ``maps'', encode motion via their phase contours, whose spacing is inversely proportional to the displacement between interactions, and whose tangent indicates the speed of motion. The fact that space-time intervals are directly encoded, not just their ratio giving the speed, indicates that systems don't evolve continuously, but only through interactions over intervals of arbitrary size.

In scalar diffraction theory, the natural metaphor for describing interactions between systems is a lens. Under various conditions, a lens performs a precise Fourier transform on an optical input. Furthermore, it only works one way, performing a Fourier transform but never an inverse Fourier transform. The inverse is achieved via the appropriate geometrical arrangement, and rather than invert the original transform, it repeats it, treating frequencies like positions and positions like frequencies. The result is the reversal of the original signal.\cite{GOODMAN2004}

\subsection{Signal evolution in this theory}
Two distinct concepts are imported from scalar diffraction theory (SDT) to help describe signal evolution in this theory. 

\subsubsection{Propagation along null intervals
\label{sec:null_intervals}}
Propagation in this theory is described analogous to a two dimensional image $f(x,y)$ traveling along the $z$-axis in SDT. In SDT the parameters of the traditional wavefunction $\Psi(x,t)$ or $\tilde{\Psi}(k,\omega)$ are evolved by Fourier transforming the incoming signal into its dual space, evolving the phase of those frequency components (there is no `time' or `space' in the frequency domain for true `propagation'), and inverting the transform back to the original domain, see eqs. $\ref{eqn:RFT_PROPAGATE_X}$ and $\ref{eqn:RFT_PROPAGATE_K}$. The end result looks like propagation, but this is misleading because no calculation was made involving the intervening space.

In this theory we begin with four independent parameters, and signal evolution is achieved similarly, without assuming time as an evolution parameter.
This choice liberates the time-variable from its one-to-one role with space: there is no meaning to the expression $x(t)$ or $y(t)$ along the path of propagation, just as there is no meaning to the expression $x(z)$ or $y(z)$ in SDT. The usual time dependency only emerges at the endpoints of the trajectory, when the system is re-expressed in the spacetime domain.

It has been understood since at least Lewis \cite{LEWIS1926-1}  that photons do exactly this as they `travel' along a light-cone, without picking up any phase evolution in the spacetime domain. Light can be measured along a trajectory through spacetime but cannot be presumed to \textit{actually be} at each specific point along that path.

The term ``null interval'' describes a Lorentz invariant spacetime interval, $\text{proper time}=\sqrt{c^2t^2-x^2}$, which is always zero. This implies that the signal is traveling at the speed of light, and is the limiting case of the Lorentz transformations for space and time, i.e. ``time dilation'' and ``length contraction'', respectively.

The mathematical result that time and space don't seem to have any bearing on the physics of light, except as measured by outside observers, is a conundrum. While their is no proper time to describe the journey between two lightlike events, time has clearly advanced at the spacetime location at which the signal is detected, relative to the emission event, when viewed from an inertial frame. Light advances time discretely in amounts equal to its duration of travel, but doesn't itself travel through the spaces or times between emission and absorption.

Importantly, the expression for phase applied in k-space in eq. \ref{eqn:RFT_PROPAGATE_K} performs a null interval translation by applying a linear phase factor in k-space. After transforming the expression into its frequency representation, one merely has to shift the phase of every frequency component appropriately to accomplish an arbitrary translation. Because the phase is invariant, there is no actual sense of time evolution, yet the system can be detected at the end of its trajectory after that phase factor is applied.

So the $k$-phase map completely describes the propagation of light (in our simplified model, excluding spin and polarization) along a null trajectory. The result is a discrete shift along an allowable path through x-space.

Just as the propagation of a photon through the space between a source and a detector (or even between the stars) is governed by a null interval with an invariant set to zero (proper time), we will use an analogous process in the dual domain to accomplish shifts in energy between states.

To see this, consider scalar diffraction theory, in which propagation in physical space (usually denoted the $z$-direction, but analogous to $t$-direction in our model) is accomplished by multiplying phase factors in the $k$-domain. By multiplying every frequency by an appropriate phase factor, time is shifted. For our photon lifecycle, because of the relation $k=\omega c$, this is a null interval, as discussed in the previous section. By symmetry, the reverse must be true: a phase factor applied to $\Psi(x,t)$ and then \textit{inverse Fourier transformed} must be a shift in the energy or momentum of the system. 


\subsubsection{Interactions are like lenses in SDT}

The second distinct concept is interaction, described by eqs. $\ref{eq:RFT_LENS_X}$ and $\ref{eq:RFT_LENS_K}$ (in contrast to eq. $\ref{eqn:RFT_PROPAGATE_X}$), which is analogous to a 2-lens optical system in SDT, comprised of repeated forward Fourier transforms. These equations can be applied to wavefunctions in either domain.

Applying \ref{eq:RFT_LENS_X} and \ref{eq:RFT_LENS_K} in succession to a distribution in X space obtains
\begin{align}
    \Psi'(-x,-t) = \mathcal{F}\left\{e^{-iS_k(k,\omega)}\mathcal{F}\left\{e^{iS_x(x,t)}\Psi\right\}\right\},
    \label{eqn:RFT_EXPLICIT_X}
\end{align}
in which the output signal is inverted.

The key feature of interaction is that it is comprised of only forward transforms, which describes real change between systems, while the forward and inverse recursive transform (which applies to propagation) is a `virtual process' with no actual effect on the system, due to the null interval discussion earlier.

In quantum optics, experiments shows that the maximum efficiency between emitter and absorber in the exchange of a photon is achieved when the temporal profile of the detector is equal to the time-reverse of the emitted wave packet.\cite{Thomas2026}\cite{Yessenov2022} \cite{wan2011time} \cite{chong2010coherent} \cite{Golla2012} \cite{STOBINSKA2009} \cite{Quabis2000} 
Thomas, et. al. report ``when the incident wave packet is the time reverse of the spontaneous emission wave packet of a target transition, the absorption probability at a receiving node can be maximized.''

Classical theory predicts both a forward and backward in time moving solution to the differential (Maxwell) equations.\cite{Leuchs2012TimeReversal} Which solution applies at emission or absorption depends on inserting boundary conditions which select, for instance, an exponentially decaying photon or an exponentially growing photon.

In the model here, the Fourier transform properties of a double lens system automatically reverses the signal and propagates it in the same action.
Goodman writes ``(In the double lens system) the output appears inverted in plane P3 due to the imaging operation, or equivalently due to the fact that a sequence of two Fourier transforms has been used, rather than one transform followed by its inverse.''\cite{GOODMAN2004}


\subsection{Distinguishing coordinates from parameters}

Treating both space and time parameters independently is possible by modeling quantum dynamics after scalar diffraction theory and holography. In a hologram, there are two distinct notions of both space and time, as well as their counterparts, spatial frequency and temporal frequency. 

The first notion, described by `parameters', is the surface of the holographic film itself, which captures the wavelike interference patterns from the source object mixed with a reference wave. If a 2-dimensional film is parameterized by $x$ and $y$, then the holographic film captures the phase at each $x,y$ combination. The film is developed so that one phase value is transparent and the opposite phase value is opaque, and thus, when a similar wavelength laser is shone on it again, it recreates the actual interference pattern that originally existed from the reference object. Note that when we look at a hologram, we don't really see the parameters: the holographic film is intended to be invisible. The space and time defined by these parameters (of integration) is unmeasurable.

The measurable quantities of space (and time) in a hologram are distinct values from the parameters, herein called `coordinates'. What we see within the hologram is the recreated image of the reference object, and the location at which we perceive it in the film is independent of the individual markings on the holographic film. We write $x_{(i)}$ and $y_{(i)}$, but to keep them distinct we call them `coordinate intervals', and label them with subscripts in parenthesis to emphasize that they have measurable values.

\subsection{Phase maps in x- and k-space}

\begin{figure}[H]
\includegraphics[width=8.0 cm]{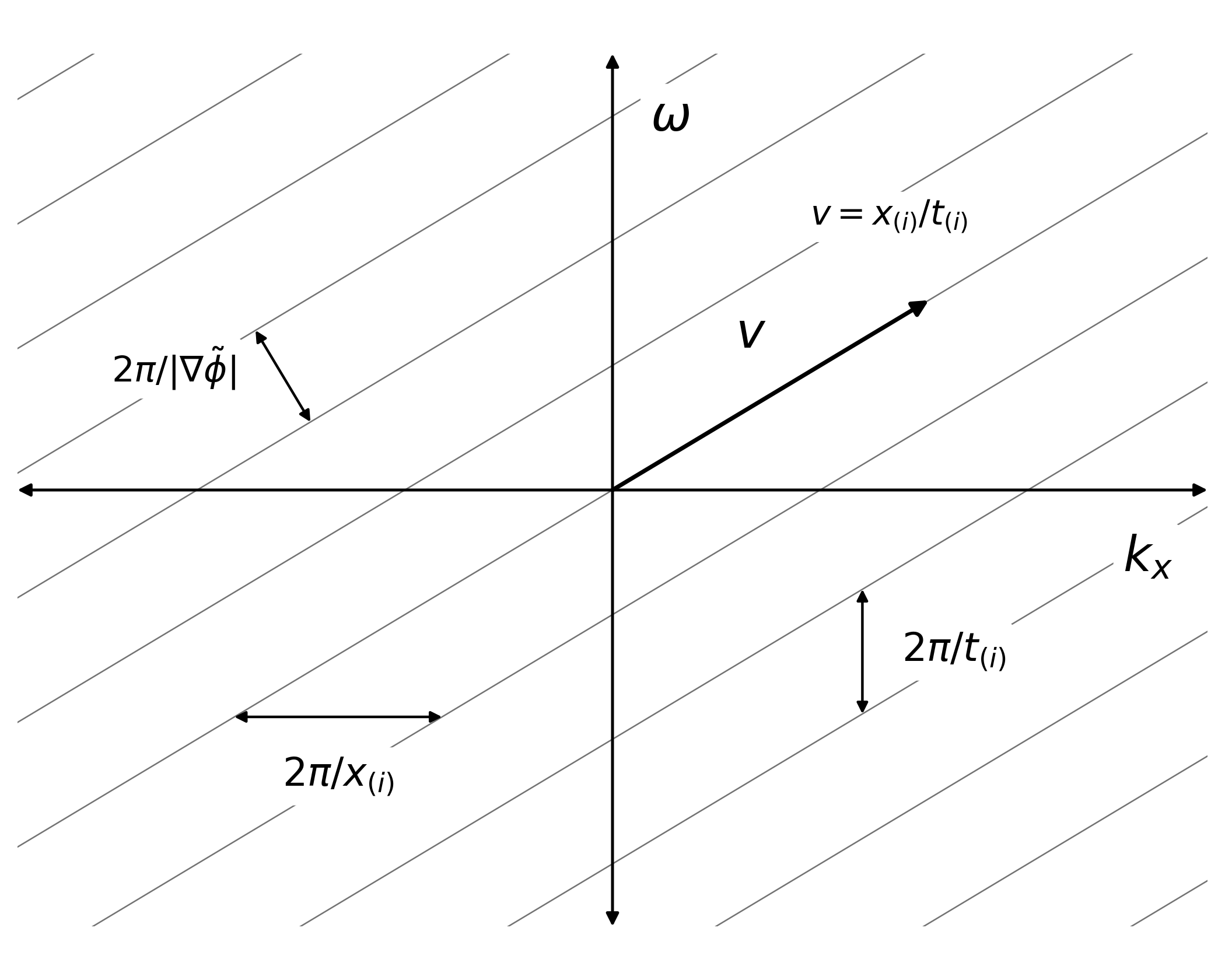}
\caption{The k-phase space contours and their relationship to familiar spacetime properties such as speed and displacement. Symbols with subscripts in parenthesis are discrete `coordinates' with measurable values. The velocity vectors are drawn for reference, plotted against spacetime parameters (not shown).\label{fig:kphase}}
\end{figure}

To understand what is meant by a ``phase map'',
imagine what information we can learn from a 2-dimensional grid in the frequency domain, parameterized by $k$ and $\omega$, the conjugates of space and time (Fig. \ref{fig:kphase}). 

For the simplest case, define a linear k-phase map
\begin{align}
    \tilde{S}(k,\omega)=kx_{(0)}-\omega t_{(0)}.
    \label{eq:kphasemap}
\end{align}
Apply eq. \ref{eq:RFT_LENS_K} to a plane wave in x-t space, $e^{i(k_{(0)}x-\omega_{(0)}t)}$, with momentum $k_{(0)}$ and energy $\omega_{(0)}$, to obtain
\begin{align}
    \begin{split}
    \label{eqn:RFT_main_eqn4}
    \Psi'(x,t) &= \mathcal{F}\{e^{-i\tilde{S}(k,\omega)} \, \mathcal{F}\{e^{i(k_{(0)}x-\omega_{(0)}t)}\}\} \\
    &= \mathcal{F}\{ e^{-i (k x_{(0)}-\omega t_{(0)}) } \, \delta(k-k_{(0)})\delta(\omega-\omega_{(0)}) \} \\
    &= e^{i(k_{(0)}(-x-x_{(0)})-\omega_{(0)}(-t-t_{(0)}))},
    \end{split}
\end{align}
that is, the wave is translated by $(x_{(0)},t_{(0)})$, the coordinates defined in the phase factor (ignoring for now the reversal in the unobservable time parameter).

For the dual case, apply the x-phase map $S(x,t)=k_{(0)}x-\omega_{(0)}t$ to a wavepacket $\tilde{\Psi}_1(k,\omega)=\tilde{A}(k,\omega)\,e^{i(kx_{(i)}-\omega t_{(i)})}$:
\begin{align}
    \begin{split}
    \label{eqn:RFT_main_eqn5}
    \tilde{\Psi}'(k,\omega) &= \mathcal{F}\{e^{-i(k_{(0)}x-\omega_{(0)}t)} \, \mathcal{F}\{\tilde{A}(k,\omega)\,e^{i(kx_{(i)}-\omega t_{(i)})}\}\} \\
    &= \mathcal{F}\{A(x-x_{(i)},t- t_{(i)})\} \\
    &= \tilde{A}(-k-k_{(0)}, -\omega-\omega_{(0)})\, e^{i((-k-k_{(0)})x_{(i)}-(-\omega-\omega_{(0)})t_{(i)})}.
    \end{split}
\end{align}

The `x-phase map' in x space is, from eq. \ref{eqn:RFT_main_eqn4}, 
\[S(x,t)=(k_{(0)}(-x-x_{(i)})-\omega_{(0)}(-t-t_{(i)}))\]
and the `k-phase map' in k space is, from eq. \ref{eqn:RFT_main_eqn5}, 
\[\tilde{S}(k,\omega)=(-k-k_{(0)})x_{(i)}-(-\omega-\omega_{(0)})t_{(i)}\]
The k-phase map $\tilde{S}(k,\omega)$, shown in fig. $\ref{fig:kphase}$, has contours of constant phase that tell us about the motion of the system. The angle of the straight contours with respect to $\omega$ and $k$ indicates the proportion of 
translation through space with respect to time, 
\[\frac{x_{(i)}}{t_{(i)}}\]
whereas the spacing between the contours is inversely proportional to the displacement 
\[\nabla \tilde{S}=x_{(i)} \hat{k}+t_{(i)}\hat{\omega}\] between the initial and final spacetime positions of the system, rather than the ratio.

We derive these values by noting that the four-dimensional distributions do not evolve in time, because time is one of the parameters. Rather, they encode evolution, so the maps themselves cannot change before and after the interaction. Comparing the phase before and after,
\begin{align}
    (k-k_{(0)})x_{(i)}-(\omega-\omega_{(0)})t_{(i)}) = kx_{(i)}-\omega t_{(i)},
\end{align}
we obtain the desired relation,
\begin{align}
    \omega_{(0)}/k_{(0)}=x_{(i)}/t_{(i)}.
\end{align}
\begin{figure}[H]
\includegraphics[width=8.0 cm]{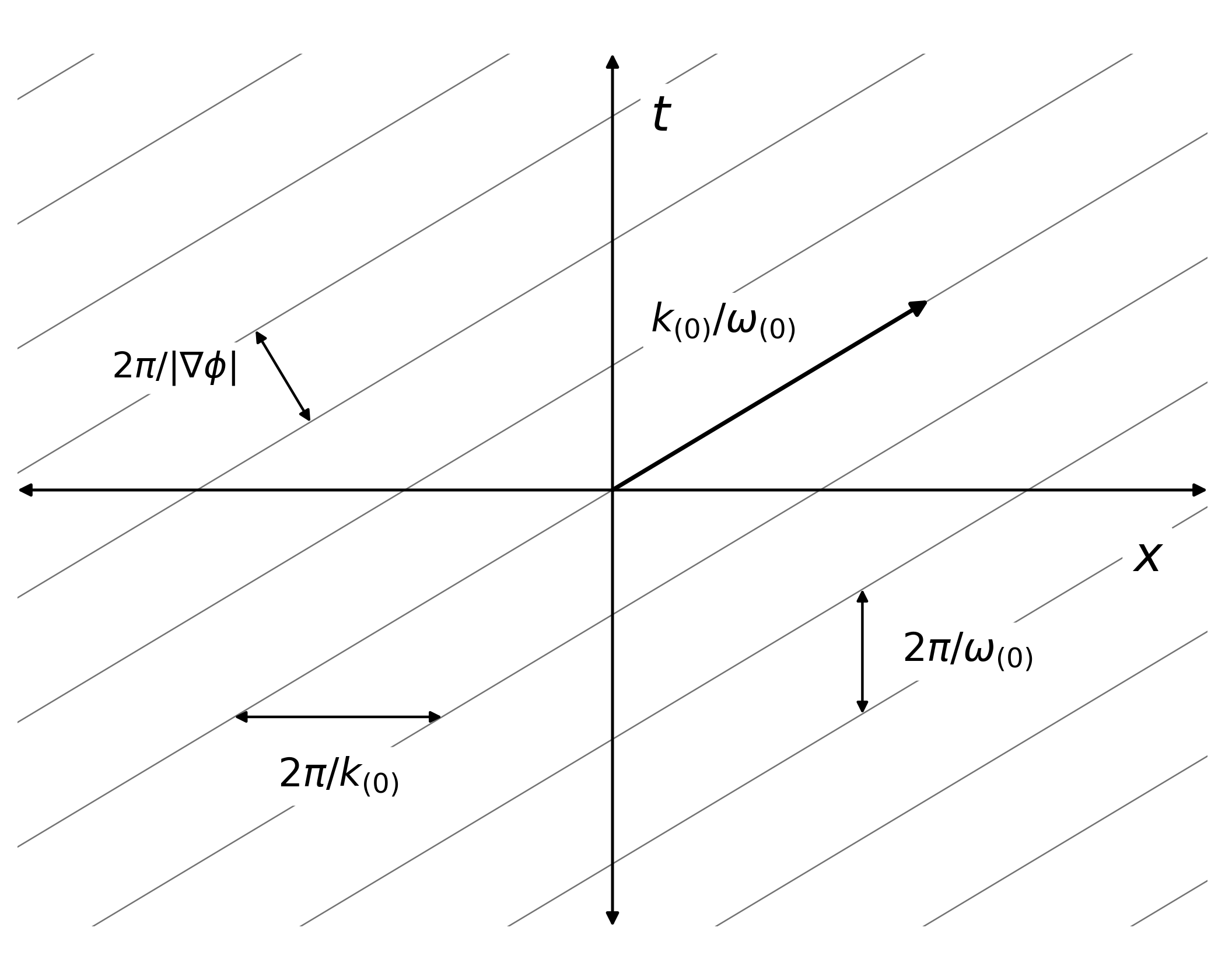}
\caption{The x-phase map has analogous properties, defined for the dual domain.\label{fig:xphase}}
\end{figure}

\section{Newton's first law: Linear phase maps encode constant motion}

As discussed in sec. $\ref{sec:digital_signals}$,
the k-phase map determines the place and time when a feature in some signal will occur. In our case, these determine where and when a photon emission or absorption event occurs. As discussed in sec. \ref{sec:null_intervals}, the x-phase map gives conjugate information about the frequency domain during these same processes, namely which wavelengths and frequencies the signal (photon) contains.

Using these two domains and their respective phase maps, we will analyze the creation, propagation, and absorption of a simplified photon.

\subsection{How long is a photon?}

\begin{figure}[H]
\includegraphics[width=12.0 cm]{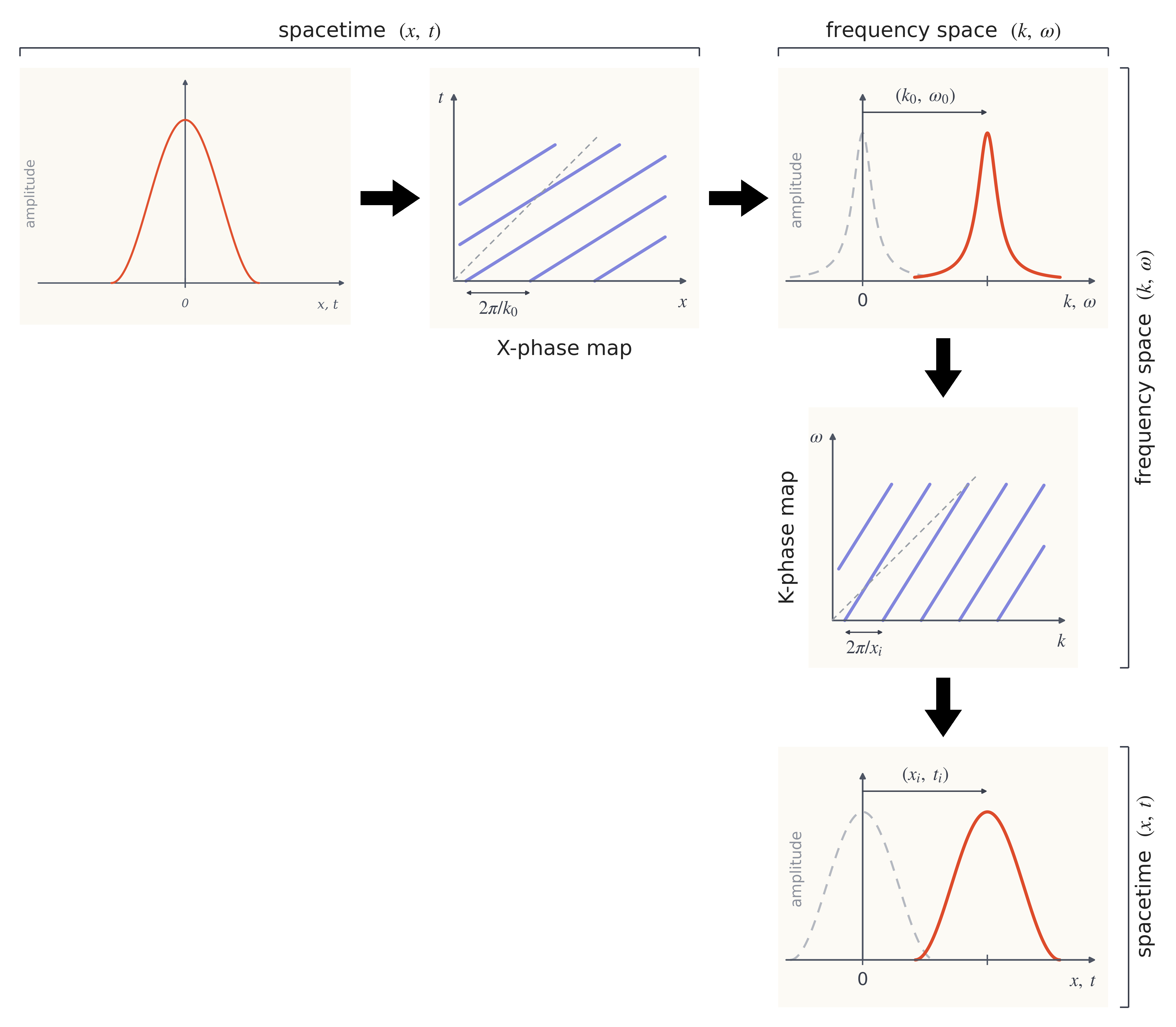}
\caption{The linear x-phase map (panel on top, middle) performs a shift in the energy spectrum of an atom, corresponding to emission of a photon. The k-phase map (panel on right, middle) performs a shift on the spacetime coordinates of the photon, propagating it.
\label{fig:lifecycle}}
\end{figure}

Fig. \ref{fig:lifecycle} shows the recursive process of alternating x- and k- domains through five panels moving right and down, whose final panel is the source of the next cycle.
We model only the spectral and spatio-temporal modes of the photon, ignoring its spin or polarization degrees of freedom. 

We assume there exists an external driving gaussian pulse $A(x,t)$ near the Rabi frequency for a given two-state system, and an atom with an available excited energy state such that the energy difference can be written
\begin{align}
    \omega_{eg}=\frac{E_e-E_g}{\hbar}.
    \end{align}
In the full quantum field theory, the creation operator $a^{\dagger}$ is used to increase the occupation number of the excited mode, and the amplitude is found using Schr\"{o}dinger's equation to be 
\begin{align}
    \int dt A(t)\, e^{i(\omega - \omega_{eg})t}.
    \label{eq:null_freq_interval}
\end{align}

This process is the first step in fig. \ref{fig:lifecycle}, where the phase map in time is simply $\omega_{eg}t$ (the factor $\exp{(i\omega t)}$ is the Fourier kernel).

Labeling the external gaussian pulse as $\Psi_1(x,t)$ and writing the x-phase map, 
\begin{align}
    S(x,t)=k_{(0)} x-\omega_{(eg)} t,
    \label{eq:xphasemap}
\end{align}
where we assumed for the photon $k_{(0)}=\omega_{(0)}/c$,
eqn. \ref{eq:RFT_LENS_X} becomes
\begin{align}
    \tilde\Psi_3=\mathcal {F}\{e^{i(k_{(0)}x-\omega_{(eg)}t)}\,A(x,t)\}=\tilde A(k-k_{(0)},\,\omega-\omega_{(eg)}),
\end{align}
generating a shifted distribution in the frequency domain, shown in panel 3 (top right). This is the energy spectrum transferred to the photon by the atomic change.

This begins to address the question in the title, ``how long is a photon?'', because the null interval in the frequency domain (i.e. the process involving inverse Fourier transform of a frequency domain signal, multiplication by a linear x-t phase map, and then a forward Fourier transform (as is effectively done in eq.  \ref{eq:null_freq_interval}), precisely defines the spectral content of the photon (a real, measurable property of the wavepacket) and an associated conjugate duration in spacetime. In other words, the temporal duration of a photon during emission and absorption is defined in terms of a null interval in the frequency domain.

Next, via direct analogy with scalar diffraction theory, we multiply the frequency-domain wave distribution by a k-phase map (eq. \ref{eq:kphasemap}) using eq. \ref{eq:RFT_LENS_K}. This phase factor advances the phases of all spectral components individually. 
The gradient 
\[\nabla \tilde{S} = x_{(i)}\hat{k} + t_{(i)}\hat{\omega}\]
determines the contour spacing $2\pi/x_{(i)}$ and $2\pi/t_{(i)}$ along the $k$ and $\omega$ directions, indicating the overall magnitude of the translation intervals in spacetime. (See fig. \ref{fig:kphase} and fig. \ref{fig:lifecycle} panel 5, right-middle.)

Finally, another Fourier transform is performed on the combined system (eq. \ref{eq:RFT_LENS_K}) to complete the detection portion of the photon's lifecycle,
\begin{align}
    \begin{split}
    \Psi_5&=\mathcal{F}\{e^{-i(kx_{(i)}-\omega t_{(i)})}\,\tilde A(k-k_{(0)},\omega-\omega_{(eg)})\}\\
    &=A(-x-x_{(i)},\,-t-t_{(i)})\;e^{i\left(k_{(0)}(-x-x_{(i)})-\omega_{(eg)}(-t-t_{(i)})\right)}.
    \label{eq:final_photon_x}
    \end{split}
\end{align}
shown in panel 5.

The final state of the photon at the detection event is offset in spacetime by $(x_{(i)}, t_{(i)})$ and reversed in parity. The signal reversal is a result of a pair of Fourier transform operations at the respective endpoints of propagation, and is consistent with the experimental result that the probability of a successful photon exchange is highest when the detected wavepacket has the \textit{reversed} temporal profile as the emitted wavepacket.\cite{Thomas2026}

Note that this process is not dispersive of gaussian wavepackets, as the Schr\"{o}dinger equation is. If $A(x,t)$ is a gaussian in space boosted by a momentum $k_{(0)}$, the Schr\"{o}dinger equation integrates over a quadratic k-phase, because $\omega=\omega(k^2)$, resulting in a wider gaussian in k-space. Here, eq. \ref{eq:final_photon_x} has restored its gaussian amplitude to the original width due to the symmetric Fourier transform in the second stage of the cycle.

\subsection{Example: Exponential decay envelope}
Stobinska et. al. demonstrate the principle of time-reversible emission and absorption envelopes with an exponential envelope. 
Given a transition probability $\tilde{A}(\omega_{eg}) \exp{(-i \omega_{eg} (t-t_0))}$, 
the photon's spectrum arriving at an absorber is given by
\begin{align}
    \tilde{A}_+(\omega)=\frac{g^*_l}{(\omega-\omega_{eg}) - i \frac{\Gamma}{2}},
\end{align}
while the absorption spectrum corresponding to the probability that the absorber is excited is
\begin{align}
    \tilde{A}_- (\omega)=\frac{g^*_l}{(\omega-\omega_{eg}) + i \frac{\Gamma}{2}}.
    \label{eq:wigner_weisskopf}
\end{align}

Inserting $\tilde{A}_+$ into eq. \ref{eq:final_photon_x} and ignoring the k-dependence which is constrained to the light cone obtains
\begin{align}
\begin{split}
    \Psi(-t)&= \mathcal{F}\left\{e^{-i(kx_{(1)}-\omega t_{(1)})}\,\tilde{A}_+(\omega)\right\}\\
    &= \Theta(t+t_{(1)})e^{-\Gamma(t+t_{(1)})/2} e^{-i \omega_{eg} (t+t_{(1)})} ,\\\
\Psi(t)&= \Theta(-t+t_{(1)})e^{\Gamma(t-t_{(1)})/2} e^{i \omega_{eg} (t-t_{(1)})} .
\end{split}
\end{align}
The spectral envelope is reflected, indicating gradual increase in probability of absorption over time, and the time parameter is inverted in parity. The sign change comes from the double Fourier transform prescribed in eq. \ref{eq:RFT_LENS_X} and \ref{eq:RFT_LENS_K}. We've landed on the correct description, including time translation of the photon from the emitter to the absorber, using only $\tilde{A}_+$.

The authors use the method of Wigner-Weisskopf to obtain both the radiation field and the corresponding absorber spectrum, different due to the sign on $\Gamma$ in the denominator. Performing an inverse Fourier transform on eq. $\ref{eq:wigner_weisskopf}$, the result is the same, but in their case they needed two different solutions to match the conditions and hand inserted the time shift $t_{(1)}$ to match the coordinates of the absorber.

\section{Newton's second law: Curved phase maps}
To expand our analysis of phase maps from the linear case to other cases, it is useful to restate a few facts. First, a linear phase map corresponds to a null interval in one domain, while corresponding to translation in the conjugate domain. Second, even a linear phase map modifies every frequency's phase, albeit in a very simple pattern (ramp) across the entire domain. It is the \textit{pattern} of phase modification, not its scope, that determines the effect on the conjugate domain. Finally, the gradient of a linear phase map is a constant which indicates a `constant velocity' in each domain, identified with the ratio of slopes in each direction. For instance, in Fig. $\ref{fig:kphase}$ the speed comes from the ratio $\frac{x_{(i)}}{t_{(i)}}$, which are each partial derivatives of the k-phase map, eq. \ref{eq:kphasemap}.

From this it seems reasonable to guess that accelerated motion might be associated with phase maps whose slope is not constant (i.e. whose gradient is not uniform). The proper machinery for capturing the second derivatives is the Hessian matrix, whose properties determine the curvature of a wavefront. Any further discussion of the Hessian is beyond the scope of this paper, but we will look at a couple basic scenarios.

The following definitions will be used,
\begin{align}
    \begin{split}
        X(k) &= \partial_k \tilde S(k,\omega)\\
        T(\omega) &= \partial_\omega \tilde S(k,\omega)\\
        K(x) &= \partial_x S(x,t)\\
        \Omega(t) &= \partial_t  S(x,t )\\
        \nabla S(x,t) &= K(x)\hat{x}-\Omega(t)\hat{t}\\
        \nabla \tilde{S}(k,\omega) &= X(k)\hat{k}-T(\omega)\hat{\omega}
        \label{eq:definitions}
    \end{split}
\end{align}
We will begin by finding phase maps that reproduce Newtonian acceleration in a straight line, then consider phase maps that will lead to simple harmonic motion, and finally, relativistic boosts and translations.

\subsection{Newtonian acceleration in a straight line}
Our goal is to identify the proper phase maps in both $k-$ and $x-$ domains which reproduce the relationship for acceleration, $x\propto t^2$.\footnote{We will always work in 1+1D for simplicity, but the extension to 4-vectors $k^\mu$ and $x^\mu$ is straightforward.} 
We will choose the phase maps so that they reproduce the expected dynamics. Alternately, if we already know both phase maps in an arbitrary situation, we can in principle evaluate it to predict the resulting dynamics.

We begin (for convenience) in the frequency domain, using eq. $\ref{eq:RFT_LENS_X}$
\begin{align}
    \Psi'(-x,-t) = \mathcal{F}\left\{e^{-i\tilde{S}(k,\omega)}\tilde{\Psi}(k,\omega) \right\},
    \label{eqn:RFT_EXPLICIT_STEP1}
\end{align}
where the negative sign on the arguments on the left is again the unusual protocol of performing a \textit{forward} Fourier transform on a \textit{frequency space} distribution.

\subsubsection{Step 2: Choose the incoming wavepacket}
Choosing a momentum-shifted gaussian for the incoming spacetime wavepacket, 
\begin{align}
    \Psi(x,t)=e^{i (k_{(0)} x-\omega_{(0)} t
    )}e^{-\frac{x^2}{2 \sigma^2}}
    \label{eq:x_gauss}
\end{align}
the frequency distribution is 
\begin{align}
    \tilde{\Psi}(k,\omega)=e^{-\frac{\sigma^2}{2} (k-k_{(0)})^2}\delta(\omega-\omega_{(0)}),
    \label{eq:k_gauss}
\end{align}
indicating an envelope of spatial frequencies centered on a carrier wavenumber $k_0$. 

\subsubsection{Choose the k-phase map and transform to x-space}
A reasonable first choice to step beyond a linear map in k-space is a map which is quadratic in $k$ only,
\begin{align*}
    \tilde{S}(k,\omega)=k^2 a_{(0)}-\omega t_{(0)},
    \label{eq:kchirp_phasemap}
\end{align*}
known as a `chirp' in signal processing. (The dimensions of $a_{(0)}$ are $[L]^{1/2}$.)

Using eq. $\ref{eq:definitions}$ will compute a partial derivative we will need momentarily,
$X(k_{(0)}) = 2 a_{(0)} k_{(0)}$.

Inserting eq. $\ref{eq:k_gauss}$ and $\ref{eq:kchirp_phasemap}$ into eq. $\ref{eqn:RFT_EXPLICIT_STEP1}$, the x-space gaussian after applying the first k-phase map becomes
\begin{align}
    \Psi'(-x,-t) &= \mathcal{F}\left\{e^{-i (k^2 a_{(0)}-\omega t_{(0)})}    e^{-\frac{\sigma^2}{2} (k-k_{(0)})^2}\delta(\omega-\omega_{(0)}) \right\}\\
    &= e^{-i \omega_{(0)}\left(t-t_{(0)}\right)}\, e^{i k_{(0)} (x- 2k_{(0)}a_{(0)})}\, e^{-\frac{(x-2 k_{(0)}a_{(0)})^{2}}{2(\sigma^{2} + 2 i a_{(0)})}},
    \label{eq:CURVED_STEP_1}
\end{align}
up to normalization. The gaussian is complex (with width $\sigma^{2} + 2 i a_{(0)}$, the time parameter is linear and shifted by $t_{(0)}$, and the $x$ parameter is linear, but the shift in $x\rightarrow x+X(k_{(0)})$ is scaled by momentum, as a result of the quadratic phase map in k.

The gaussian envelope now has an imaginary part, 
\begin{align}
    \exp{\left(i\frac{D}{2}(x-2k_{(0)}a_{(0)})^2\right)},
    \label{eq:chirp}
\end{align}
where 
\begin{align}
    D\equiv \frac{2a_{(0)}}{\sigma^4+4a_{(0)}^2}.
\end{align}
As a result, we identify a contribution to the x-phase map,
\begin{align}
   S(x,t)= \frac{D}{2}(x-2k_{(0)}a_{(0)})^2+k_{(0)}x-\omega_{(0)}t+\text{constant terms}.
   \label{eq:x-phase-accel-n2}
\end{align}
This x-phase term, quadratic in x, is introduced into $\Psi(x,t)$, as a result of executing the Fourier integral over a quadratic phase in the dual domain.
In other words, the introduction of chirp in the k-phase map carries over to x-space.

\subsubsection{Choose the x-phase map and transform to k-space}
To generate acceleration in space, we want an asymmetric relationship between space and time, so it seems reasonable to keep the time/frequency domains linear, as in the previous example. Thus our x-phase factor for this integration cycle is (see eq. \ref{eq:xphasemap})
\begin{align}
    S_1(x,t) = k_{(\Delta)} x - \omega_{(\Delta)} t
    \label{eq:newton_x_phase}
\end{align}
where $k_{(\Delta)}$ and $\omega_{(\Delta)}$ are finite shifts in momentum and energy.

However, the chirp or quadratic phase in $x$ (eq. $\ref{eq:x-phase-accel-n2}$) introduced into the wavefunction via the k-phase dependence leads to precisely the changing gradient that we postulated in order to describe accelerated motion through x-space for a wave packet. Considering both the x-phase from eq. \ref{eq:x-phase-accel-n2} resulting from the previous Fourier transform, and the term in eq. $\ref{eq:newton_x_phase}$ introduced each cycle, the slope of the x-phase map becomes
\begin{align}
    \partial_{x} \left(S(x,t)+S_1(x,t)\right) = D (x -2 k_{(0)}a_{(0)})+k_{(0)}+k_{(\Delta)}.
    \label{eq:newton_xphase_slope}
\end{align}
It is always the slope of the phase in any given direction which generates a shift in the conjugate variable under a Fourier transform. The \textit{linear} shift in k resulting from eq. $\ref{eq:newton_xphase_slope}$ is  $k_{(0)}+k_{(\Delta)}$, and after n cycles it will be 
\begin{align}
    k_{(0)}+nk_{(\Delta)}.
    \label{eq:accel_n_kick}
\end{align}
Additionally, there is a dispersive term proportional to $x-2 k_{(0)} a_{(0)}$. This partially cancels the dispersion from the previous Fourier transform step, and it generates another gaussian with a conjugate chirp the next time we visit the k-phase map, which is perpetuated each cycle. 

These cyclical updates happen each time the full cycle of transforms is applied, and lead to a shift in position which is quadratic to the time steps, $t_{(0)}$, describing a wave packet whose center is accelerating.

Each successive cycle the \textit{change} in x, which is given generally by the value of $X(k)=\partial_k \tilde{S}(k,\omega)$, gets uniformly larger; $2 a_{(0)} k_{(\Delta)}$ gets added to the the \textit{change} each cycle. By contrast, the time coordinates advance linearly as $t\rightarrow t+t_{(0)}$ in each cycle.

Thus, non-relativistic, Newtonian kinematics can be reproduced with the phase maps 
\begin{align}
    \begin{split}
        \tilde{S}(k,\omega)&=ak^2+\omega t_{(0)}\\
        S(x,t)&=k_{(\Delta)}x-\omega_{(\Delta)}t.
        \label{eq:phase_maps_accel}
    \end{split}
\end{align}
As with the Schr\"{o}dinger equation, the components of the gaussian disperse as it propagates.

We have kept in mind the distinction between coordinates (with parenthesis subscripts) and parameters (without). It is important not to attribute any fixed or measurable value to the parameters, but it is allowable to \textit{shift} them by a known coordinate. It is the progressive shifts (extracted from the slope of the phase maps) that correspond to motion, not the values of the parameters or phase maps themselves.

\subsection{One-dimensional harmonic motion}
In the case of linear acceleration, the phase maps in eq. \ref{eq:phase_maps_accel} were linear in $S(x)$ but quadratic in $\tilde{S}(k)$. This kicked the momentum during each application of two Fourier transforms, and the position was always further increased by this increasing value, leading to increasing velocity in a single direction. 

If instead we make the $S(x)$ dependence also quadratic, the momentum $k$ receives kicks in both directions gradually over time, so the system replicates a spring force (Hooke's Law). The resulting motion is harmonic.

\subsubsection{Identify the phase maps}
Harmonic motion involves forces, so the results in the previous section indicate that the quadratic $k$-dependence in the k-phase map (eq. $\ref{eq:kphasemap}$) is necessary. To try one further step away from the inertial motion condition (linear phase maps), we will try modifying the x-phase map, eq. $\ref{eq:xphasemap}$, to be quadratic in $x$,
\begin{align}
    S_1(x,t) = -\frac{\m}{2}x^2 - \omega_{(\Delta)} t,
    \label{eq:xphasemap_harmonic}.
\end{align}
Note that the coefficient to the quadratic is now negative. This will have the effect of making the shift in $k$ always opposite to the displacement, like a restoring force.

During the transformation from k- to x-space, as before, the wavepacket is displaced to $2k_{(0)}a_{(0)}$ and a quadratic phase factor is contributed, as in eq. \ref{eq:chirp}.

\subsubsection{Apply the x-phase map and transform to k-space}
Adding eq. \ref{eq:xphasemap_harmonic} to eq. \ref{eq:x-phase-accel-n2}, the slope is
\begin{align}
    \partial_x\left(S+S_1\right) = (D-\m)\left(x-2k_{(0)}a_{(0)}\right) + k_{(0)} - 2\m k_{(0)}a_{(0)}.
\end{align}
In addition to the shift in x, there is also a shift in k, $k_{(0)} \rightarrow k_{(0)} - 2\m k_{(0)}a_{(0)}$, is now proportional to the negative of the displacement, a restoring force.

The phase maps in simple harmonic motion reflect the potential of simple harmonic motion in the usual case. It is evident that the phase maps are closely related to the potential energy of a system.

The Hermite-Gauss functions used here are eigen functions of the Fourier transform and form a closed set, so every action stays within the group.

\section{Demonstrating Lorentz invariance}

\subsection{Relativistic boost}
The model presented here captures all dynamics through maps that exist in the phases of 3+1D `wave distributions'. Propagations are passive and are accomplished via recursions of forward and inverse Fourier transform calculations with a particular phase factor in each domain, modeled after diffractive optics theory. By contrast, interactions are active and are accomplished by forward Fourier transforms alone.

To adequately describe displacements in spacetime that are consistent with special relativity, this method must,  through its operation, produce intervals that undergo proper Lorentz transformation. This is almost trivially guaranteed since the phase is well-known to be relativistically invariant. In practice, this is true because wavenumber and frequency form a Lorentz transform pair just as space and time do, and the combination $kx-\omega t$ is an invariant under boosts.

Importantly, the Lorentz transform is change in perspective not a physical process, so we rely on the propagation model (i.e. alternating forward and inverse transforms) rather than the lens model (forward-only Fourier transforms).

To demonstrate Lorentz invariance of the wave distributions in 3+1D space, define three `cross-domain' phase maps,
\begin{align}
\begin{split}
    S_{1} &= a k t \\
    S_{2} &= b \omega x \\
    S_{3} &= S_{1},
    \label{eq:LORENTZ_PHASE_MAPS}
\end{split}
\end{align}

We propose to implement the propagation equation eq. \ref{eqn:RFT_PROPAGATE_X} separately in the position and time parameters, $x$ and $t$, in the following order:
\begin{align}
    \Psi_1 &= \mathcal{F}_\text{k}^{-1}\!\left\{ e^{-i S_1}\, \mathcal{F}_\text{x}\!\left\{ \Psi_0  \right\} \right\} \\
    \Psi_2 &= \mathcal{F}_{\omega}^{-1}\!\left\{ e^{i S_2}\, \mathcal{F}_\text{t}\!\left\{ \Psi_1 \right\} \right\} \\
    \Psi_3 &= \mathcal{F}_\text{k}^{-1}\!\left\{ e^{-i S_1}\, \mathcal{F}_\text{x}\!\left\{ \Psi_2 \right\} \right\}
    \label{eq:RFT_LORENTZ}
\end{align}

Starting from a moving gaussian wavepacket at fixed $k_{(0)}$ and $\omega_{(0)}$, insert eq. \ref{eq:LORENTZ_PHASE_MAPS} into eq. \ref{eq:RFT_LORENTZ}, omitting the real gaussian amplitude $A(x,t)$ which is unaffected by the linear phase maps. 

\textbf{Step 1:}
\begin{align}
\begin{split}
    \Psi_1(x',t)&=\mathcal{F}_{\text{k}}^{-1}\!\left\{ e^{-iakt}\,\mathcal{F}_{\text{x}}\!\left\{ A(x,t)e^{i(k_0 x - \omega_0 t)} \right\} \right\}\\
    &= \int dk\, e^{ikx'}\, e^{-iakt}\int dx\, e^{-ikx}\, e^{i(k_0 x - \omega_0 t)} \\
    &= \int dk\, e^{ik(x' - at)}\, e^{-i\omega_0 t}\,\delta(k - k_0) \\
    &= e^{i\left(k_0 x' - (\omega_0 + a k_0)t\right)}
\end{split}
\end{align}

\textbf{Step 2:}
Insert previous result into the next recursive Fourier transform over time and frequency domain.

\begin{align}
\begin{split}
    \Psi_2(x',t')&= \mathcal{F}_{\omega}^{-1}\Big\{ e^{ib\omega x'} \mathcal{F}_{\text{t}}\big\{ e^{-i\omega_{(0)} t} e^{ik_{(0)}(x' - a t)} \big\} \Big\} \\
    &= \int d\omega \, e^{-i\omega t'} e^{ib\omega x'} \int dt \, e^{i\omega t} \Big( e^{i(k_{(0)} x' - (\omega_{(0)} + a k_{(0)}) t)} \Big) \\
    &= e^{ik_{(0)} x'} \int d\omega \, e^{i\omega(b x' - t')} \, \delta\big(\omega - (\omega_{(0)} + a k_{(0)})\big) \\
    &= \exp\Big(i\big(((1 + ab) k_{(0)} + b \omega_{(0)}) x' - (\omega_{(0)} + a k_{(0)}) t'\big)\Big)
\end{split}
\end{align}

\textbf{Step 3:}
Insert the result from Step 2 into a recursive Fourier transform over spatial domain again.
\begin{align}
\begin{split}
    \Psi_3(x'',t') &=\mathcal{F}^{-1}_{\text{k}}\{ e^{-i a k t'}\, \mathcal{F}_{\text{x}}\{\Psi_2(x', t')\}\} \\
    &= \int\! dk\, e^{ik x''}\, e^{-iak t'}\,e^{-i\left(\omega_{(0)} + a\,k_{(0)}\right)t'}\int\! dx'\, e^{-ik x'}\,e^{i\left((1+ab)\,k_{(0)} + b\,\omega_{(0)}\right)x'} \\
    &= e^{i\left[(1+ab)\,k_{(0)} + b\,\omega_{(0)}\right]x''}\;    e^{-i\left[(1+ab)\,\omega_{(0)} + a(2+ab)\,k_{(0)}\right]t'} 
\end{split}
\end{align}
Over the three steps, the x-phase map went from 
\begin{align}
    k_{(0)} x + \omega_{(0)} t \rightarrow \left((1+ab)\,k_{(0)} + b\,\omega_{(0)}\right)x''-((1+ab)\,\omega_{(0)} + a(2+ab)\,k_{(0)})t'
    \label{eq:LORENTZ_XPHASE_MAP_COMPARE_ab}
\end{align}
Next, define rapidity $\eta$ such that the relativistic speed $\beta = \tanh(\eta)$, and set
\begin{align}
\begin{split}
    \qquad a \equiv \tanh\!\left(\tfrac{\eta}{2}\right)\\ 
    \quad b \equiv \sinh(\eta) \\
    1 + ab = \cosh(\eta)
\end{split}
\end{align}
With these definitions, and using $\sinh\eta = a(1+\cosh\eta) = a(2+ab)$, eq. \ref{eq:LORENTZ_XPHASE_MAP_COMPARE_ab} can be written
\begin{align}
    k_{(0)} x - \omega_{(0)} t = \Big( cosh(\eta) k_{(0)} + sinh(\eta) \omega_{(0)}\Big) x - \Big( sinh(\eta) k_{(0)} + cosh(\eta) \omega_{(0)}\Big) t
    \label{eq:LORENTZ_XPHASE_MAP_COMPARE_sinh}
\end{align}

To interpret this, consider that, unlike the Schr\"{o}dinger wavefunction, the 3+1D distributions here are not \textit{functions} of time, i.e. they cannot evolve from `before' to `after', so we can simply write $\Psi_3(x,t) = \Psi(x,t)$, above. Eq. \ref{eq:LORENTZ_XPHASE_MAP_COMPARE_ab} becomes
\begin{align}
    k_{(0)} \rightarrow k_{(0)}cosh(\eta)  + \omega_{(0)}sinh(\eta)  \\
    \omega_{(0)} \rightarrow k_{(0)}sinh(\eta)  + \omega_{(0)}cosh(\eta) 
\end{align}
Thus, the momentum and energy coordinates in one frame get rotated into a combination of momentum and energy in the new frame. This is a standard result (in hyperbolic form) for the Lorentz transform, which can be manipulated into the standard form by substituting $\beta = \frac{v}{c} \equiv tanh(\eta)$, and $\gamma = \frac{1}{\sqrt{1-\beta^2}} \equiv cosh(\eta) $.

Thus, the process of applying phase factors in the frequency domain to advance a signal both in space and in time is relativistically invariant. This is a strong clue that the phase maps defined in this way, drawn from quantum theory, but consistent with special relativistic constraints, are an important or even fundamental description of dynamical motion in spacetime.

\subsection{Relativistic inertial motion}
As a final exploration, we will evaluate eqs. \ref{eq:RFT_LENS_X} and \ref{eq:RFT_LENS_K} for timelike propagation using a phase map based on the invariant interval in Minkowski space.

\subsubsection{Identify the phase maps and define the incoming wavepacket}
Using the dispersion relations from special relativity, the phase maps are 
\begin{align}
    S(x,t) &= m_{(0)} \sqrt{t^2-x^2}\\
    \tilde{S}(k,\omega) &= -\tau_{(0)} \sqrt{\omega^2-k^2}.
    \label{eq:xk_phase_map_rel_propagation}
\end{align}

We will again choose a gaussian envelope, this time in both $x$ and $t$, since Lorentz transforms will rotate $x$ and $t$ into each other, so time cannot have an asymmetric form from space.

Anticipating Minkowski geometry, we define hyperbolic variables,
\begin{align}
    \begin{split}
        x &= \rho \sinh(\xi)\\
        t &= \rho \cosh(\xi)\\
        k &= \mu \sinh(\theta)\\
        \omega &= \mu \cosh(\theta),
        \label{eq:hyperbolic_defs_1}
    \end{split}
\end{align}
where $\theta$ and $\xi$ are rapidities, a hyperbolic alternative to velocity $\beta$. This simplifies the radical in the phase map and the invariant interval discussed in the previous section on Lorentz invariance, $kx-\omega t$,
\begin{align}
    \begin{split}
        \sqrt{t^2 - x^2} &= \rho \\    
        \sqrt{\omega^2 - k^2} &= \mu \\
        kx-\omega t &= -\mu\rho  \cosh(\xi - \theta)
        \label{eq:hyperbolic_defs_2}
    \end{split}    
\end{align}

In this coordinate system, the gaussian wave packet in the frequency domain can be written as
\begin{align}
    \tilde{\Psi}(\mu,\theta)=e^{-\frac{1}{2 \Delta^2}(\mu-m_{(0)})^2}\,\delta(\theta-\theta_{(0)}),
    \label{eq: gaussian_hyperbolic}
\end{align}
where $\Delta$ is the standard deviation, and the measure for the Fourier integral becomes $dk d\omega = \mu d\mu d\theta$, so the Fourier transform can be written in hyperbolic coordinates as
\begin{align}
    \Psi'(-\rho,-\xi) &= \int_0^{\infty}\mu\,d\mu\!\int\! d\theta\;
    e^{-i\mu\left(\rho\cosh(\xi-\theta)-\tau_{(0)}\right)}\,
    e^{-\frac{(\mu-m_{(0)})^2}{2\Delta^2}}\,\delta(\theta-\theta_{(0)})\\
     &= \int_0^{\infty}\mu\,d\mu
    \,e^{-i\mu s(\rho,\xi)}\, e^{-\frac{(\mu-m_{(0)})^2}{2\Delta^2}},
    \label{eq:hankel_transform}
\end{align}
a Hankel/Fourier-Helgason transform. We have labeled $s(\rho,\xi) \equiv \rho  \cosh(\xi - \theta_{(0)}) - \tau_{(0)}$.
The integral over $\theta$ was easily performed, and the integral over $\mu$ is in the class of gaussian derivatives, readily integrable using integration by parts, resulting in
\begin{align}
    \Psi'(s) &= \sqrt{2\pi}\,\Delta\,\left(m_{(0)} - i\Delta^{2} s\right) e^{-\frac{1}{2}\Delta^{2}s^{2}}\, e^{-i m_{(0)} s}.
\end{align}
where, at the center of the packet, $s \equiv \rho \cosh{(\xi - \theta_{(0)})} - \tau_{(0)}$ can also be written in terms of the linear coordinates, $-(k_{(0)}x-\omega_{(0)} t)/m_{(0)} - \tau_{(0)}$.

We have already seen that the phase quantity $k_{(0)} x - \omega_{(0)} t$ is invariant under Lorentz transforms, so we have shown that $\Psi$ depends on x and t only through the invariant $s$, so the displacement it encodes is the same in every frame. Here, the center of the wavepacket was displaced by the proper time $\tau_{(0)}$.

Using eqs. \ref{eqn:RFT_PROPAGATE_X} and \ref{eqn:RFT_PROPAGATE_K} together with the phase maps (eq. \ref{eq:xk_phase_map_rel_propagation}), we obtained inertial motion in special relativity. 
Invariance under $\omega$-$k$ space propagation follows the same steps in the dual spaces. 


\section{Conclusion}
Defining what is meant by the temporal duration of a photon presents a challenge because the emission and absorption occur in a different domain than the propagation. Modeling wave front propagation as a recursive Fourier transform process provides a basic framework through which this can be understood and is a simple method which has numerous features, including being a quantum theory that is relativistically invariant. It does so by treating time equivalently to space, as independent parameters whose phase map encodes motion.


\bibliography{references} 


\end{document}